\documentclass[letterpaper, 10 pt, conference]{ieeeconf}  
\IEEEoverridecommandlockouts
\usepackage{amsmath,amssymb,amsfonts}
\usepackage{bm}
\usepackage{algorithm}
\usepackage{algpseudocode}
\usepackage{graphicx}
\usepackage{textcomp}
\usepackage{xcolor}
\usepackage[normalem]{ulem}
\usepackage{subcaption}
\usepackage{xspace}
\usepackage{microtype}
\usepackage{tikz} 

\def\BibTeX{{\rm B\kern-.05em{\sc i\kern-.025em b}\kern-.08em
    T\kern-.1667em\lower.7ex\hbox{E}\kern-.125emX}}
\def\ri#1{\textcolor{blue}{RI: #1}}
\def\nh#1{\textcolor{red}{NH: #1}}
\def\bl#1{\textcolor{orange}{BL: #1}}
\def\lb#1{\textcolor{teal}{LB: #1}}
\def\ri#1{}
\def\nh#1{}
\def\bl#1{}
\def\lb#1{}
\def\code#1{{\footnotesize\texttt{#1}}} 

\newtheorem{theorem}{Theorem}
\newtheorem{lemma}{Lemma}
\newtheorem{problem}{Problem}
\newtheorem{example}{Example}
\newtheorem{definition}{Definition}

\def\timeDomHigh#1{[#1]^\uparrow}

\def\rotogo{\rho^\looparrowright}
\newcommand{\rotogotext}{Ro-To-Go\xspace}

\algnewcommand\algorithmicswitch{\textbf{switch}}
\algnewcommand\algorithmiccase{\textbf{case}}
\algnewcommand\algorithmicassert{\texttt{assert}}
\algnewcommand\Assert[1]{\State \algorithmicassert(#1)}%
\algdef{SE}[SWITCH]{Switch}{EndSwitch}[1]{\algorithmicswitch\ #1\ \algorithmicdo}{\algorithmicend\ \algorithmicswitch}%
\algdef{SE}[CASE]{Case}{EndCase}[1]{\algorithmiccase\ #1}{\algorithmicend\ \algorithmiccase}%
\algtext*{EndSwitch}%
\algtext*{EndCase}%

\algdef{SE}[DOWHILE]{Do}{doWhile}{\algorithmicdo}[1]{\algorithmicwhile\ #1}%

\newcommand\copyrighttext{%
  \footnotesize \textcopyright \the\year{} IEEE. Personal use of this material is permitted. Permission from IEEE must be obtained for all other uses, including reprinting/republishing this material for advertising or promotional purposes, collecting new collected works for resale or redistribution to servers or lists, or reuse of any copyrighted component of this work in other works.}

\newcommand\copyrightnotice{%
\begin{tikzpicture}[remember picture,overlay]
\node[anchor=south,yshift=10pt] at (current page.south) {\fbox{\parbox{\dimexpr0.75\textwidth-\fboxsep-\fboxrule\relax}{\copyrighttext}}};
\end{tikzpicture}%
}

\title{\LARGE \bf
\rotogotext!\\Robust Reactive Control with Signal Temporal Logic
}
\author{Roland Ilyes, Lara Bruderm{\"u}ller, Nick Hawes, Bruno Lacerda
\thanks{This work received EPSRC funding via the "From Sensing to Collaboration" programme grant [EP/V000748/1]. R.I. and L.B. were supported by Amazon Web Services Lighthouse scholarships. } 
\thanks{All authors are with the Oxford Robotics Institute, University of Oxford, United Kingdom. For correspondence: {\tt\footnotesize\{rolandilyes, larab, nickh, bruno\}@robots.ox.ac.uk}}
}

\begin{document}
\maketitle
\copyrightnotice
\thispagestyle{empty}
\pagestyle{empty}

\begin{abstract}
Signal Temporal Logic robustness is a common objective for optimal robot control, but its dependence on history limits the robot's decision-making capabilities when used in model predictive control approaches. In this work, we introduce Signal Temporal Logic robustness-to-go, a new quantitative semantics for the logic that isolates the contributions of suffix trajectories. We prove its relationship to formula progression for Metric Temporal Logic, and show that the robustness-to-go depends only on the suffix trajectory and progressed formula. We implement robustness-to-go as the objective in a model predictive control algorithm\lb{this only makes sense in an model predictive control setting anyways right?}\ri{not necessarily, for example one can use it to ignore portions of a trajectory in post-hoc analysis for debugging purposes} and use formula progression to efficiently evaluate it online. We test the algorithm in simulation and compare it to model predictive control using other robustness measures. Our experiments show that using robustness-to-go improves performance compared to using traditional robustness. 


\end{abstract}


\section{Introduction}
\label{sec:Introduction}


Many robotic applications demand complex behaviors that are difficult to specify. Temporal Logics \cite{TL} provide a grounded mathematical approach to specify such behaviors. Linear Temporal Logic (LTL) \cite{LTL} is a popular logic for discrete-time robotic applications. In recent years, there has been a growing interest in logics that describe more complex behaviors than LTL. Metric Temporal Logic (MTL) \cite{MTL} and Metric Interval Temporal Logic (MITL) \cite{MITL} extend LTL to reason about real-valued time. Signal Temporal Logic (STL) \cite{STL} is a subset of MITL defined over real-valued signals, making it particularly useful to robotic applications. In addition to providing a binary answer for whether a signal satisfies a specification, it also admits a quantitative score of how well the signal satisfies it, known as the \textit{Robust Satisfaction Value}, or simply \textit{robustness} \cite{STLRobustness}. This value characterizes how robust a planned trajectory is to unforeseen disturbances. 

The state-of-the-art robot control algorithms for STL specifications are based on trajectory optimization\ri{cite?}\lb{will you?} techniques, and are split into two distinct approaches. One approach is to encode the STL specification as mixed integer constraints on an optimization problem. For convex objective functions, this results in a mixed integer convex program (MICP) \cite{MILP1}. Although this approach is complete, the constraints induced by the STL specification introduce new variables for each time step and this scales poorly. This approach also requires both the system dynamics \cite{smooth4} and the logical predicates \cite{MILP2} to be linear. The other approach directly optimizes the STL robustness \cite{smooth2}, resulting in a nonlinear programming problem (NLP). This approach allows for nonlinear system dynamics and logical predicates. Although incomplete, this approach allows for nonlinear system dynamics and logical predicates, and scales better than the MICP approach due to the lack of integer constraints. 


\begin{figure}[t]
  \begin{center}
    \includegraphics[trim = {0cm, 0cm, 0cm, 0cm},clip,width=0.95\linewidth]{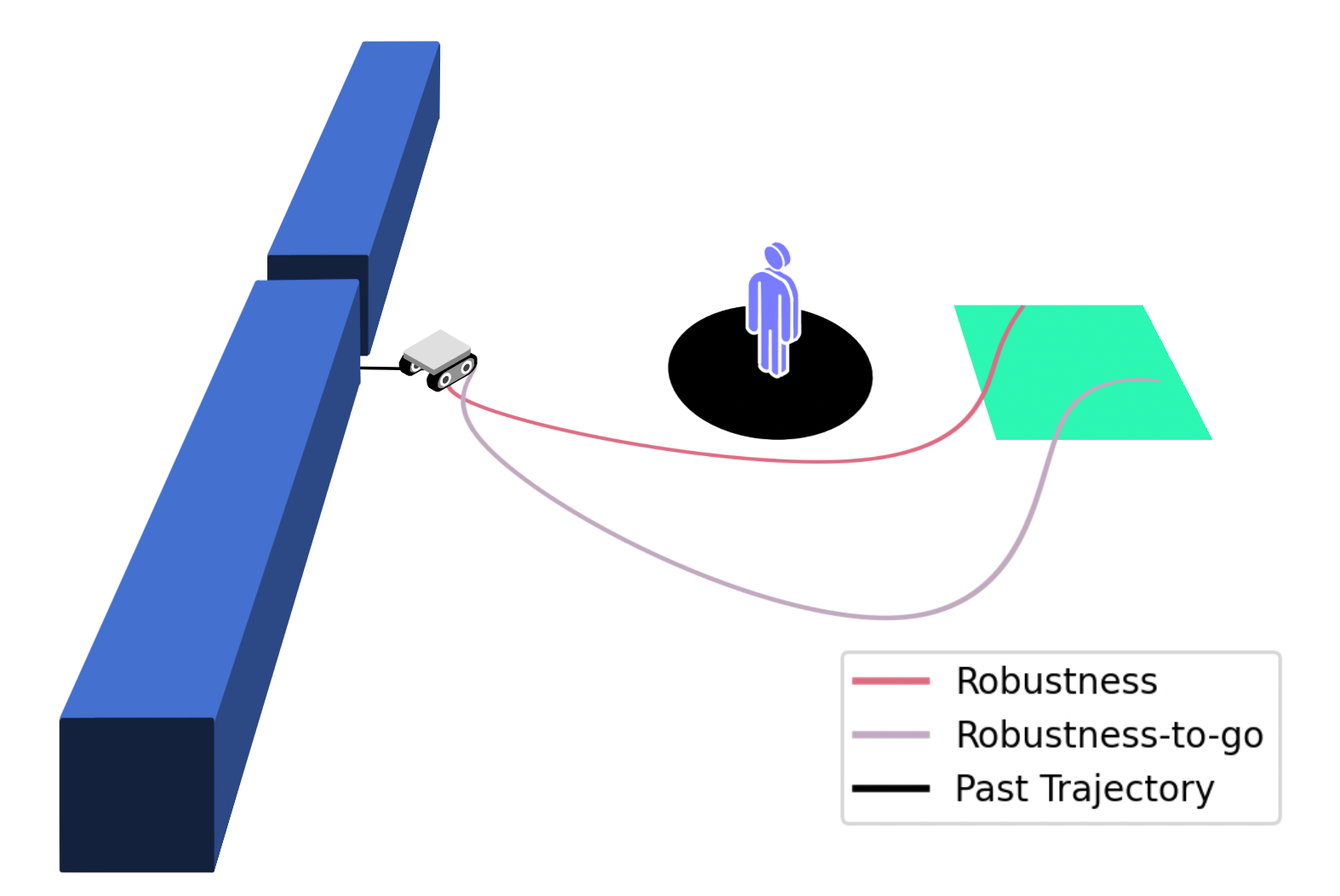}
  \end{center}
  \caption{
  This examples shows a robot planning a path to the goal while avoiding the human. With regular robustness, the proximity to the obstacles at the beginning limits the overall robustness, so there is no value in going further away from the human. With robustness-to-go, the robot has forgotten about its past proximity to the obstacles and can find trajectories that are more robust to human movement.}
\label{fig: ex1}
\end{figure}

 
In addition to planning for complex behaviors, robots must also react to unforeseen changes in their environment. Consider a practical example, shown in Fig.~\ref{fig: ex1}, where a robot must reach a goal within a fixed time limit while avoiding obstacles and humans. 
For such a mission, the robot might be unaware of how the human will move and must react to the human's motion.

One approach to achieve such\lb{kind of} reactive behavior is Model Predictive Control (MPC), which solves the aforementioned optimization problems periodically during mission execution. Given the current state, at each iteration, a new optimal trajectory is computed and executed until the next trajectory is found. This iterative replanning gives MPC approaches some level of resilience\bl{why resilient instead of robust?} to dynamic environments. 

While the STL robustness is a sound objective function for open-loop control synthesis, it has shortcomings in the context of feedback control through MPC. This is because the robustness is a function of the entire trajectory, and so the entire history of the robot's motion must be considered in the objective function.
If the robot comes close to violating the specification, then all of its plans from that point onward are limited in how robust they can be. After such an event, the robot cannot distinguish between good and bad trajectories because the robustness does not provide any incentive for the robot to move further away from future possible violations.
Fig. \ref{fig: ex1} captures this phenomenon. Both trajectories have the same robustness due to their proximity to obstacles at the beginning, even though the blue\ri{? colorblidn} trajectory steers further from the human.


    In this paper, we introduce \textit{robustness-to-go} (\rotogotext), a new robustness measure designed to reason about the \textit{suffix} of trajectories, the portion beyond a given time. This measure scores the trajectory from the given time onwards, instead of scoring the past trajectory of the robot. This measure is recursively defined on the syntax of the logic and generalizes to all of STL. At each MPC planning step, we use \rotogotext as our objective function, enabling the robot to remain robust to future disturbances, regardless of its past. We prove that \rotogotext can be efficiently calculated by using MITL formula progression \cite{formProg}, a technique that rewrites formulas iteratively after each observation.


The contributions of this paper are (i) introduction of robustness-to-go, a new STL robustness measure that focuses on reasoning about suffix trajectories, (ii) a technique to efficiently calculate the robustness-to-go using formula progression (iii) proofs of the soundness of the measure and its relationship to formula progression, and (iv) a series of case studies investigating the incorporation of this measure in a feedback control algorithm, and its impact on robot behavior. 

\section{Related Work}
\label{sec:RelatedWork}

Control synthesis for STL specifications was first investigated in \cite{MILP1}, where they extend an MICP-approach to synthesis for LTL specifications from \cite{MILPLTL} to handle STL specifications. This approach automatically generates a set of mixed integer linear constraints for an optimal control problem. If the system is linear and the objective function is convex, then the optimal solution can be found using a MICP \cite{MILP2}. 

These encodings only apply to STL formulas over linear predicates. Furthermore, nonlinear systems turn the optimization problem into a mixed-integer non-convex program, which is intractable. Therefore, instead of encoding the STL specification as mixed integer constraints, an alternate approach formulates an optimization problem where the specification is instead considered in the objective function, via the robustness measure. This approach still results in a nonconvex optimization problem, but finding local optima is faster because of the lack of integer constraints. The notion of optimizing temporal logic robustness was first introduced in \cite{nonSmooth}, in the context of minimizing MTL robustness for verification purposes. Robustness is a highly discontinuous function, and so the authors in \cite{smooth2} propose a smooth approximation of MTL robustness, and use gradient ascent tools to maximize it for control synthesis purposes. The works \cite{smooth3}, \cite{smooth5}, \cite{smooth6} extend this approach to optimize smooth approximations of STL robustness.

Robustness as an objective function, however, has its shortcomings. Called \textit{locality} in \cite{daddyBelta}, traditional robustness characterizes trajectories by their \textit{most critical point}, that is, the point that requires the smallest alteration to change the satisfaction result. In the context of open-loop control synthesis, this is a reasonable objective to optimize for: making the most critical point more robust to disturbances makes the entire trajectory more robust to disturbances. In online feedback control, however, the most critical point could be a point in the past, that has already happened. The result of this is that all plans will be limited by this point, and the robot has trouble identifying trajectories that are more robust \textit{from this point onwards}. Works \cite{smooth4} and \cite{smooth1} attempt to better characterize the entire trajectory by defining new ``average'' STL robustness measures to optimize for. Recently, the authors in \cite{daddyBelta} generalize these robustness measures, among others, by introducing \textit{Generalized Mean Robustness}. These measures are not subject to the locality effect, meaning that the robot can distinguish between two trajectories whose most critical points are the same. In contrast, this work addresses the issue of locality specifically in the context of online control, where the most critical point happened in the robot's past. 
We propose \textit{robustness-to-go}, a measure that is defined on the \textit{suffix} of trajectories. While still local, the most critical point for a formula's robustness-to-go is within the planning horizon of the robot. 

\section{Preliminaries}
\label{sec:Preliminaries}
Let $\bm{x} \in X \subseteq \mathbb{R}^n$ be a \textit{state}, and $\mathcal{T}$ be a set of time instants $t$ such that $\mathcal{T} \subseteq \mathbb{R}_{\geq 0}$. Then, a \textit{signal} $s$ is a mapping from $\mathcal{T}$ to $X$, and \textit{sample} from the signal is $(t,s(t))$. 

A \textit{signal predicate} is a formula of the form $g(\bm{x})>0$, where $g: X \rightarrow \mathbb{R}$. We express properties of signals with respect to signal predicates using \textit{signal temporal logic} (STL) \cite{STL}.

\begin{definition}
  (STL Syntax). \textit{
The STL Syntax is recursively defined by:
\begin{equation*}
  \varphi := \top \ | \ g(\bm{x}) > 0 \ | \ \lnot \varphi \ | \ \varphi \land \varphi \ | \ \varphi U_I \varphi
\end{equation*}
where 
$\varphi$ are STL formulas and $I = \langle a,b \rangle$ is a time interval that can be open or closed on both ends with $0 \leq a < b \leq \infty$. Notation $\top$, $\lnot$, and $\land$ are Boolean ``true'', ``negation'' and ``conjunction,'' respectively, and U denotes the temporal ``until'' operator.
  }
\end{definition}

Boolean disjunction can be expressed as $\varphi \lor \psi = \lnot (\lnot \varphi \land \lnot \psi)$, Boolean implication as $\varphi \rightarrow \psi = (\lnot \varphi) \lor \psi$, and Boolean ``false'' as $\bot = \lnot \top$\bl{Let's axe this portion, these are usual connections}. The temporal operator \textit{eventually} ($\Diamond_I \varphi$) can be expressed as $\top U_I \varphi$ and the temporal operator \textit{globally} ($\square_I \varphi$) can be expressed as $\lnot \Diamond_I \lnot \varphi$. The interval $I+c = \langle a+c, b+c \rangle$.


\begin{definition}
  (STL Semantics). \textit{
The semantics of STL is defined over a signal s at time t as:
\begin{align*}
  & (s,t) \vDash \top && \iff \top \\
   & (s,t) \vDash g(\bm{x})>0 && \iff g(s(t)) > 0 \\
   & (s,t) \vDash \lnot \varphi&& \iff (s,t) \nvDash \varphi \\
   & (s,t) \vDash \varphi \land \psi && \iff (s,t) \vDash \varphi \land (s,t) \vDash \psi \\
   & (s,t) \vDash \varphi U_I \psi && \iff \exists t' \in I+t \ s.t. 
   (s,t') \vDash \psi \\ 
   & && \quad \quad \ \ \land \forall t'' \in [t, t'), (s,t'') \vDash \varphi\\
\end{align*}
where $\vDash$ denotes satisfaction. A signal s satisfies an STL formula $\varphi$ if $(s,0) \vDash \varphi$.
  }
  \label{def: stlSemantics}
\end{definition}

In this work, we consider the set $\mathcal{T}$ to be finite and strictly increasing, and employ discrete-time, or \textit{pointwise}, semantics. Refer to \cite{handbook} and \cite{ctFromDt} for a discussion on the differences between continuous and pointwise semantics.

The authors in \cite{formProg} show that, under pointwise semantics, we can incrementally track the satisfaction of MTL formulas using \textit{formula progression}. We adapt it here for STL.

\begin{definition}
  (STL Formula Progression). \textit{
    Given an STL formula $\varphi$, a time step $\Delta$, and an observation of a state $x$, the progressed formula is given by $P(\varphi,\Delta,\bm{x})$}:
\begin{alignat*}{2}
  &P(\top,\Delta,\bm{x}) &&= \top \\
  &P(\bot,\Delta,\bm{x}) &&= \bot \\
  &P(g(\bm{x})>0,\Delta,\bm{x}) &&= 
  \begin{cases}
    \top \text{ if } g(\bm{x}) > 0 \\
    \bot \text{ otherwise }
  \end{cases}
  \\
  &P(\lnot \varphi,\Delta,\bm{x}) &&= \lnot P(\varphi,\Delta,\bm{x})\\
  &P(\varphi \land \psi,\Delta,\bm{x}) &&= P(\varphi,\Delta,\bm{x}) \land P(\psi,\Delta,\bm{x})\\
  &P(\varphi U_I \psi,\Delta,\bm{x}) &&= 
  \begin{cases}
    P(\varphi,\Delta,\bm{x}) \land \varphi U_{I^\leftarrow_\Delta} \psi &\text{if } 0 < I\\
    P(\psi,\Delta,\bm{x}) \lor  &\\
    \ \ (P(\varphi,\Delta,\bm{x}) \land \varphi U_{I^\leftarrow_\Delta} \psi) & \text{if } 0 \in I,
  \end{cases}
\end{alignat*}
where $I^\leftarrow_\Delta = (I-\Delta) \cap \mathbb{R}_{\geq 0}$.
  \label{def: formProg}
\end{definition}

Note that we have added a truncation operation to the definition from \cite{formProg} to enforce nonnegative interval endpoints. Here, $I^\leftarrow_\Delta$ is the subset of $I-\Delta$ that is greater than or equal to zero, which could be the empty set $\emptyset$. By Definition \ref{def: stlSemantics}, $\varphi U_\emptyset \psi = \bot$ because there exists no $t' \in \emptyset$. Therefore, the truncation operation resolves the case $0>I$ from \cite{formProg}, which is why it is absent from Definition \ref{def: formProg}.

The use of signal predicates allows interpretation of STL using \textit{quantitative} semantics, called \textit{robustness} \cite{STLRobustness}, in addition to qualitative semantics.

\begin{definition}
  (STL Robustness). \textit{
    The quantitative semantics of STL is defined over a signal s at time t as $\rho(s,t,\varphi)$:
\begin{alignat*}{2}
   &\rho(s,t,\top) &&= \infty \\
   &\rho(s,t,g(\bm{x})>0) &&= g(s(t)) \\
   &\rho(s,t,\lnot \varphi) &&= -\rho(s,t,\varphi) \\
   &\rho(s,t, \varphi \land \psi) &&= \min\{\rho(s,t,\varphi),\rho(s,t,\psi)\} \\
   &\rho(s,t,\varphi U_I \psi) &&= \sup_{t' \in I+t }\min
   \begin{pmatrix}
   \rho(s,t',\psi), \\
   \inf\limits_{t'' \in [t, t')}{\rho(s,t'',\varphi)} \\
   \end{pmatrix}.
\end{alignat*}
A signal s satisfies an STL formula $\varphi$ if $\rho(s,0,\varphi) > 0$.
  }
  \label{def: robustness}
\end{definition}

\bl{``this need some unpacking in words'' - Bruno}

\begin{definition}
  (STL Time Horizon \cite{vasileRipped}). \textit{
    The Time Horizon of an STL formula $\timeDomHigh{\varphi}$ is the horizon within which a signal can impact the satisfaction of $\varphi$:
  \begin{align*}
    & \timeDomHigh{g(\bm{x})>0} &&= 0 \\ 
    & \timeDomHigh{\lnot \varphi} &&= \timeDomHigh{\varphi} \\ 
    & \timeDomHigh{\varphi \land \psi} &&= \max\{\timeDomHigh{\varphi}, \timeDomHigh{\psi}\}\\ 
    & \timeDomHigh{\varphi U_{\langle a,b \rangle} \psi} &&= b + \max\{\timeDomHigh{\varphi},\timeDomHigh{\psi}\}.\\ 
  \end{align*}
  \ri{say something about the lack of a domain for truth?}
  A formula is bounded if $\timeDomHigh{\varphi} < \infty$, or unbounded otherwise.
  }
  \label{def: horizon}
\end{definition}

\section{Problem Formulation}
\label{sec:ProblemFormulation}
Consider a discrete time robotic system of the form
\begin{equation*}
  \bm{x}^r_{t+\Delta} = f(\bm{x}^r_t,\bm{u}_t)
\end{equation*}
where $\bm{x}^r\in X^r \subset \mathbb{R}^{n_r}$ is the state of the robot, $\bm{u}\in U \subset \mathbb{R}^m$ is the control input, and $\Delta \in \mathbb{R}_{\geq 0}$. The environment state is $\bm{x}^e \in X^e \subset \mathbb{R}^{n_e}$, and its dynamics are unknown. The environment can capture dynamic yet not controllable aspects of the world that are relevant to the STL formula. This can include not only a dynamic physical environment, but also other agents acting in the same space as the robot. 


We wish to specify tasks as \textit{bounded} STL formula over the composed robot-environment system. For a specification $\varphi$, we consider time domain $\mathcal{T} = \{0, \Delta, 2\Delta, ..., \timeDomHigh{\varphi}\}$. Our signal $s \in S$ is a mapping $s: \mathcal{T} \rightarrow X^r \times X^e \times U$. We denote the components of $s$ using superscripts, i.e., the robot state component of signal $s$ at time $t$ is $s^r(t)$.

We first present the motion planning problem. Given that the robot finds motion plans iteratively for closed-loop feedback control, it will find motions \textit{during} execution. Because STL satisfaction depends on the robot's future \textit{and} past, the robot must consider its history. Let $s_t$ be a signal defined over $[0,t]$ such that $t < \infty$. $s_t$ is a \textit{prefix} of $s$ if, for all $t' \leq t, s(t') = s_t(t')$.

\begin{problem}
  Given system $f$, bounded STL task $\varphi$, and signal prefix $s_t$, find $s$ such that: 
  \begin{align*}
    &\text{(i) } s \vDash \varphi, \\
    &\text{(ii) } s(t') = s_t(t') \ \forall \ t' \in [0,t], \\
    &\text{(iii) } s^e(t') = s^e_t(t') \ \forall \ t' \in [t, \timeDomHigh{\varphi}], \\
    &\text{(iv) } s^r(t' + \Delta) = f(s^r(t'),s^u(t')) \ \forall \ t' \in [0, \timeDomHigh{\varphi}]. \\
  \end{align*}

  \label{problem: motionPlanning}
\end{problem}
\noindent Here, (i) ensures the solution satisfies the STL formula, (ii) enforces the solution to contain the observed history of the system trajectory, (iii) assumes the environment is static for planning purposes, and (iv) ensures that the solution obeys the robot dynamics $f$. Due to the static environment assumption, there are no guarantees on the safety of the closed-loop controller. However, by integrating solutions to Problem \ref{problem: motionPlanning} in an MPC loop, subsequent solutions plan for the updated environment state, allowing the robot to react to the unknown environment disturbances.  


\section{\rotogotext!}
\label{sec:roToGo}

Solutions to Problem \ref{problem: motionPlanning} include both the NLP and MICP approaches discussed in section \ref{sec:Introduction}. The MICP approach introduces extra constraints to ensure $s \vDash \varphi$, and optimize for some other convex objective function \cite{MILP1}. The NLP approach involves optimizing the robustness, or some other robustness measure \cite{daddyBelta}. They find $\max_{s \in S} \rho(s,0,\varphi)$, and $s \vDash \varphi$ if the optimization converges on a local optima such that $\rho(s,0,\varphi) > 0$.  

We are interested in the NLP solution to Problem \ref{problem: motionPlanning} in the context of MPC. The robustness depends on the entire trajectory, meaning that the robot must consider the past as a \textit{prefix} $s_t$, and the robot only plans for the \textit{suffix}. As the MPC algorithm iterates and the robot executes the mission, the portion of the trajectory that the robot can control shrinks. In many instances, the portion of the trajectory in the past, outside of the robot's control, dominates the robustness. The impact of this is that the portion of the trajectory that the robot \textit{can} control does not have a big impact on the robustness, and the robot cannot easily distinguish between different behaviors.

Ideally, the robot should focus on what it \textit{can} control, and find the most robust behaviors \textit{within that window}. To this end, we propose \textit{robustness-to-go} (\rotogotext).


\begin{definition}
  (STL Robustness-To-Go). \textit{
    The robustness-to-go from time $\hat{t}$ of STL formula $\varphi$ is defined over a signal s at time t as $\rotogo(s,t,\hat{t},\varphi)$:
\begin{alignat*}{2}
  &\rotogo(s,t,\hat{t},\top) &&= \infty \\
   &\rotogo(s,t,\hat{t},g(\bm{x})\!>\!0) &&=
   \begin{cases}
     g(s(t)) & \text{ if } t > \hat{t} \\
     \infty\cdot \text{sign}(g(s(t))) & \text{ otherwise} \\
   \end{cases} \\
   &\rotogo(s,t,\hat{t},\lnot \varphi) &&= -\rotogo(s,t,\hat{t},\varphi) \\
   &\rotogo(s,t,\hat{t}, \varphi \land \psi) &&= \min\{\rotogo(s,t,\hat{t},\varphi),\rotogo(s,t,\hat{t},\psi)\} \\
   &\rotogo(s,t,\hat{t},\varphi U_I \psi) &&= \sup_{t' \in I+t}\min
   \begin{pmatrix}
     \rotogo(s,t',\hat{t},\psi), \\
     \inf\limits_{t'' \in [t, t')}{\rotogo(s,t'',\hat{t},\varphi)} \\
   \end{pmatrix}.
\end{alignat*}
  }
  \label{def: rotogo}
\end{definition}

  \rotogotext differs from traditional robustness in how it treats predicates. When the predicate robustness is evaluated after some user-specified $\hat{t}$, it is equivalent to traditional robustness. If, however, the predicate robustness is evaluated \textit{before} $\hat{t}$, the robustness is instead $\infty$ or $-\infty$, depending on if the predicate is satisfied or not. 

  We propose solving Problem \ref{problem: motionPlanning} by optimizing for the \rotogotext from the planning time onwards. This approach is sound because \rotogotext itself is a sound robustness measure.

\begin{theorem}
  (Soundness). \textit{
    A signal s satisfies an STL formula if and only if it has a positive robustness-to-go, i.e. $(s,t) \vDash \varphi \iff \rotogo(s,t,\hat{t},\varphi) > 0$.}
  \label{thm: RoToGoSound}
\end{theorem}

  \proof{The proof is in Appendix \ref{sec: proofs}.}

\begin{example}
  (Robustness-to-go.) Consider the formula $\varphi_{reachAvoid} = \square_{[0,20]} (\code{workspace} \land \lnot (\code{human}\lor \code{obs}_1 \lor \code{obs}_2)) \land \Diamond_{[15,20]} \code{goal}$, which specifies that the robot must avoid the human and obstacles for twenty seconds, and reach the goal between fifteen and twenty seconds from the start time. Here, the subformulas are defined as
\begin{align*}
  &\code{workspace} = (\bm{x}^r[0]\!>\!0 \land \bm{x}^r[0]\!<\!5 \land \bm{x}^r[1] \!>\! 0 \land \bm{x}^r[1] \!<\! 5)\\
  &\code{human} = (\bm{x}^r[0]-\bm{x}^e[0])^2 + (\bm{x}^r[1]-\bm{x}^e[1])^2 \!<\! 0.25\\
  &\code{obs}_1 = (\bm{x}^r[0]\!>\!0.5 \land \bm{x}^r[0]\!<\!1 \land \bm{x}^r[1] \!>\! 0 \land \bm{x}^r[1] \!<\! 2.4)\\
  &\code{obs}_2 = (\bm{x}^r[0]\!>\!0.5 \land \bm{x}^r[0]\!<\!1 \land \bm{x}^r[1] \!>\! 2.6 \land \bm{x}^r[1] \!<\! 5)\\
  &\code{goal} =  (\bm{x}^r[0] \!>\! 4 \land \bm{x}^r[0] \!<\! 5 \land \bm{x}^r[1] \!>\! 2 \land \bm{x}^r[1] \!<\! 3). \\ 
\end{align*}
  Fig. \ref{fig: ex1} shows the robot planning for these specifications online. The robot has already executed a prefix trajectory, and began close to $\text{obs}_1$ and $\text{obs}_2$, at $\bm{x}^r_0 = [0.5,2.5]^T$. Because of this, all trajectories beginning from this initial state will have a robustness of \textit{at most} 0.1. Indeed, the red line in Fig. \ref{fig: ex1} shows the robot coming within 0.13 meters of the human. This is because, when using robustness, the robot could not distinguish between this trajectory and one that goes further from the human. In contrast, the \rotogotext from the planning point onwards does not consider the contribution of the initial point, but only the contribution from the points within the plan under consideration. As such, it can better distinguish between good and bad motion plans, finding a plan that only comes within 0.79 meters of the human. The plan found using \rotogotext is safer with respect to unforeseen human motion.
  \label{ex: roToGo}
\end{example}

\section{Formula Progression for STL \rotogotext}

  \rotogotext is a function of the entire trajectory signal. As such, when planning trajectories during mission execution, the history of the robot's motion must be considered. During planning, these iterative \rotogotext evaluations must consider the same past. Because the robot cannot change the past, processing this data is redundant. In this section, we show how to use formula progression from Definition \ref{def: formProg} to avoid these redundant operations. Informally, we show that \textbf{the \rotogotext} from time $\hat{t}$ is the same as \textbf{the robustness of the progressed formula} at time $\hat{t}$.

\begin{theorem}
Let $\varphi_0$ be an STL formula, $s$ be a signal, and consider sample $(t_i,s(t_i))$. Let $\Phi = \{\varphi_0, \varphi, ..., \varphi_{i+1}\}$ be a set of STL formulas such that $\varphi_{k} = P(\varphi_{k-1},t_k-t_{k-1},s(t_{k-1}))$ for each $0 < k \leq i+1$. Then, 
\begin{equation*}
  t_i \geq t_0 \rightarrow \rho(s,t_{i+1},\varphi_{i+1}) = \rotogo(s,t_0,t_i,\varphi_0).
\end{equation*}
  \label{thm: progForRotogo}
\end{theorem}
  \proof{The proof is in Appendix \ref{sec: proofs}.}


\begin{figure}
    \centering
    \includegraphics[trim = {4cm, 5cm, 5cm, 18.8cm},clip,width=\linewidth]{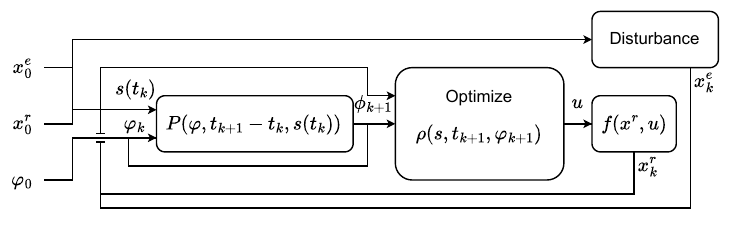}
    \caption{Control Block Diagram for \rotogotext MPC with Formula Progression.}
    \label{fig:diagram}
\end{figure}

Fig. \ref{fig:diagram} shows how to incorporate formula progression for MPC with \rotogotext. The approach begins by using the initial robot state, environment state, and STL formula $\varphi_0$ to find the progressed formula $\varphi_{k+1}$. In the NLP optimization problem, we optimize the robustness of trajectories with respect to $\varphi_{k+1}$, beginning from the next time step $t_{k+1}$. According to Theorem \ref{thm: progForRotogo}, the optimization objective $\rho(s,t_{k+1},\varphi_{k+1})$ is the \rotogotext of the original formula \textit{from the initial point onwards} $\rotogo(s,t_0,t_0,\varphi_0)$. Note that, even though the first point evaluated is $t_{k+1}$, the state at $t_k$ must still be considered to ensure that trajectories obey the robot dynamics $f$. Once the optimal solution is found, the robot applies control input $u$, and the environment evolves according to its unknown motion. The next iteration begins, with the progressed formula $\varphi_{k+1}$ becoming $\varphi_,$, and the new robot and environment states become $s(t_k)$. For a given planning iteration at time $t_i$, the robot is optimizing the \rotogotext of the original formula \textit{from the current time onwards} $\rotogo(s,t_0,t_i,\varphi_0)$.

Using formula progression to find the \rotogotext has three benefits. The first is that it avoids reprocessing of past observations. This is because the robustness of the progressed formula depends only on points within the robot's planning horizon. The second benefit is that, by reducing the problem of calculating \rotogotext to the problem of calculating robustness, we can exploit readily available tools for calculating robustness such as Breach \cite{breach} or STLRom \cite{stlrom}. Finally, using formula progression paves the road for combining \rotogotext with other robustness measures, such as those discussed in \cite{daddyBelta}.

\section{Experiments}
\label{sec:CaseStudies}


While the approach described in Fig. \ref{fig:diagram} is planner agnostic, we have implemented \rotogotext for use in solving Problem \ref{problem: motionPlanning} using VP-STO \cite{vpsto}, a state-of-the-art motion planning and MPC algorithm which uses CMA-ES \cite{cmaes} as the optimization technique. CMA-ES is zero-order optimization method and thus can handle discontinuous objective functions, making it a good fit for STL robustness.\lb{sentence too long?} To calculate the STL robustness of progressed formulas while planning, we use STLRom \cite{stlrom}. The hyperparameters for our planner are provided in Appendix \ref{sec: params}.

For our robot dynamics, we consider a two-dimensional double integrator, i.e., $\bm{x}_t^r = [p_x, p_y, v_x, v_y]^T$, $\bm{u}_t = [a_x, a_y]^T$, and
\begin{align*}
  \bm{x}_{t+\Delta}^r \! =\!  f(\bm{x}_t^r,\bm{u}_t) \! = \!    
  \begin{bmatrix} 
    \bm{I}_{2x2} & \Delta \bm{I}_{2x2} \\
    \bm{0}_{2x2} & \bm{I}_{2x2} \\
  \end{bmatrix} 
  \bm{x}_t^r\! +\!  
  \begin{bmatrix} 
    0.5 \Delta^2\bm{I}_{2x2} \\
    \bm{I}_{2x2} \\
  \end{bmatrix}
  \bm{u}_t,
\end{align*}
  where $\Delta=0.1$ is the sampling time. The robot is subject to velocity constraints of 2 m/s and input constraints of 2 m/s$^2$ in both dimensions. The environment state $\bm{x}^e \in \mathbb{R}^2$ receives a random perturbation $w = [w_1,w_2]^T$, where  $w_1,w_2\sim \mathcal{N}(0,1)$, at every planning step.

\begin{figure}
\centering
\begin{subfigure}{.5\linewidth}
  \centering
    \includegraphics[trim = {2cm, 0cm, 2cm, 1cm},clip,width=\linewidth]{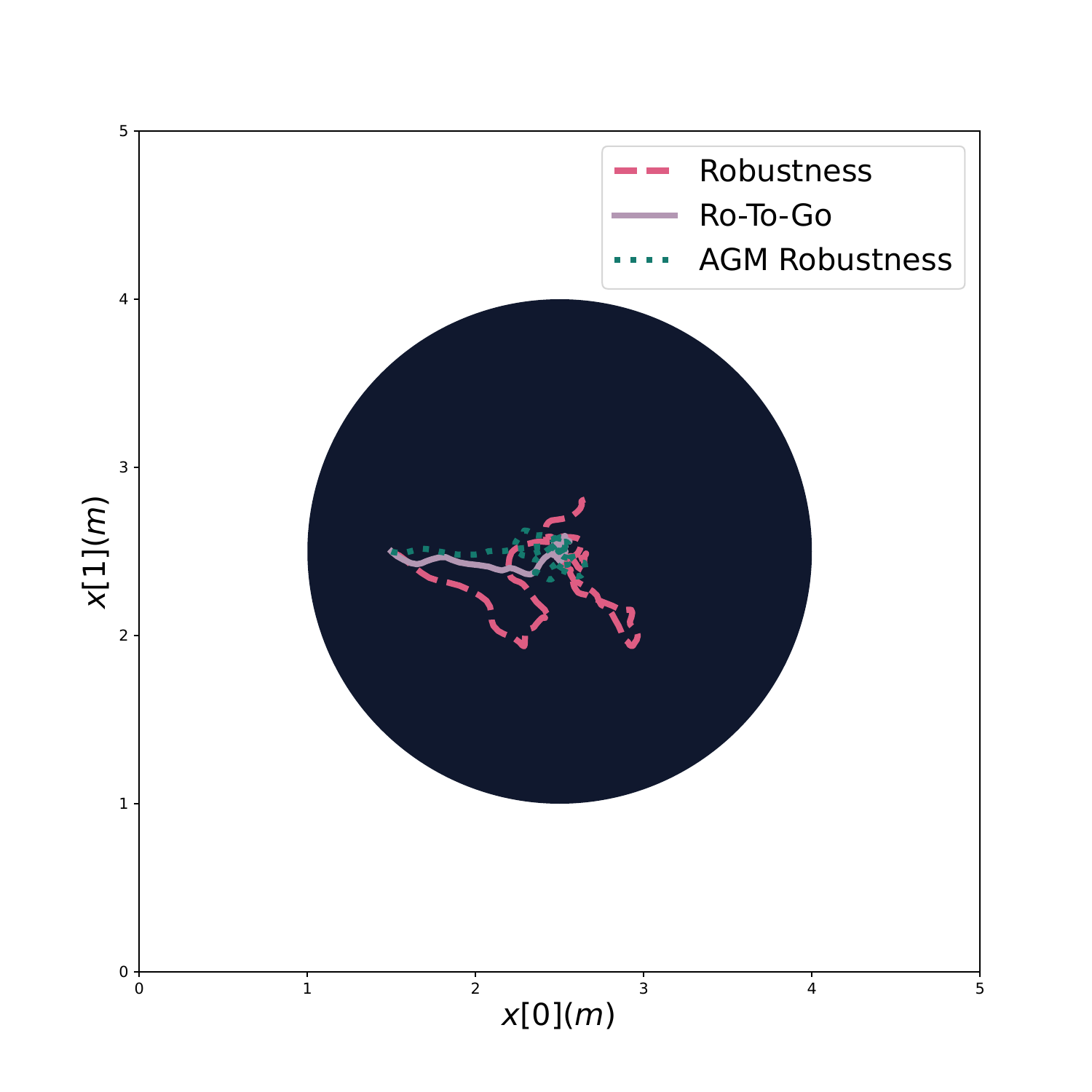}
  \caption{$\varphi_{stayIn}$}
    \label{fig:stayIn}
\end{subfigure}%
\begin{subfigure}{.5\linewidth}
  \centering
    \includegraphics[trim = {2cm, 0cm, 2cm, 1cm},clip,width=\linewidth]{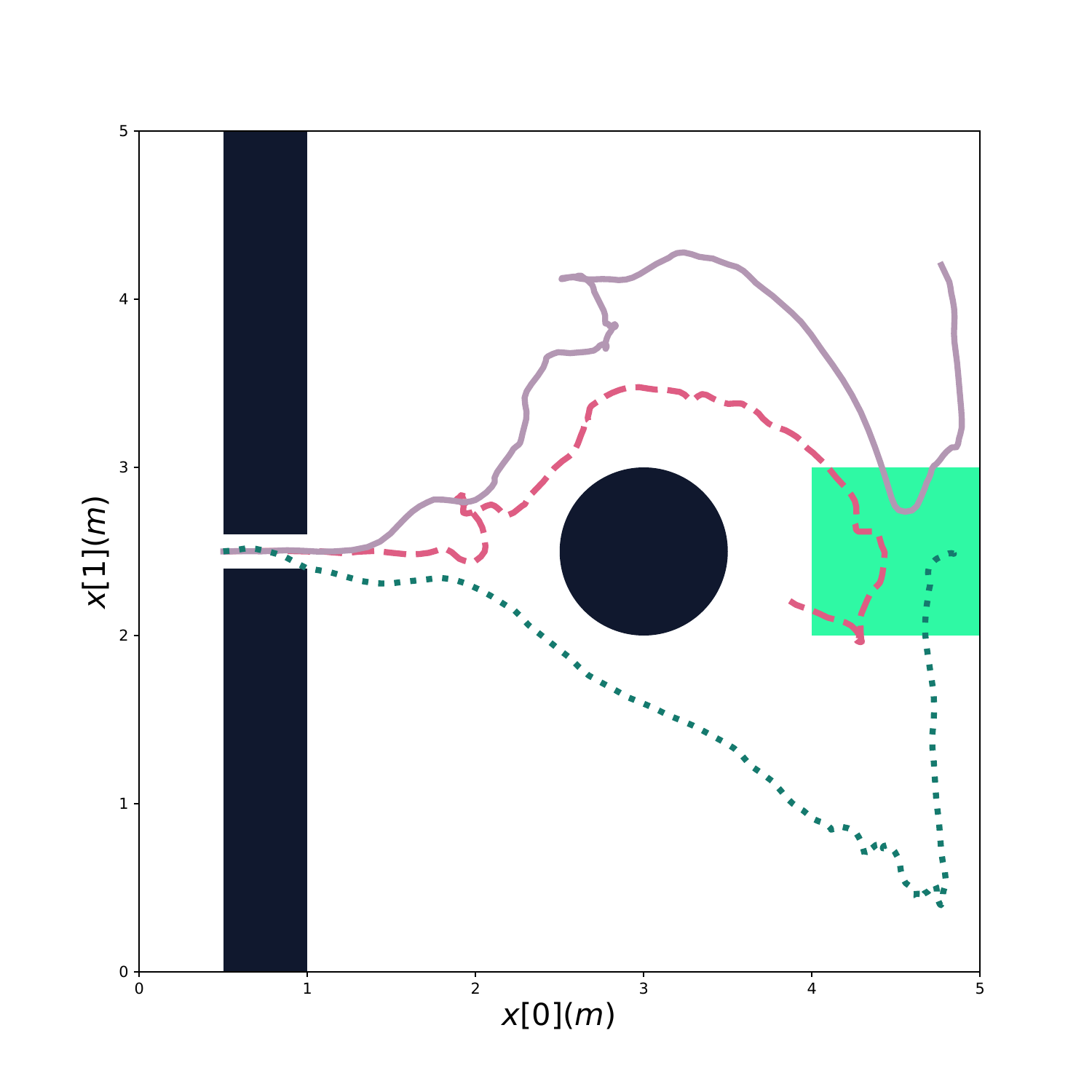}
  \caption{$\varphi_{reachAvoid}$}
    \label{fig:reachAvoid}
\end{subfigure}

\caption{Sample trajectories for two STL formulas, simulated with a static environment.}
\label{fig:bothEnvs}
\end{figure}

  For our experiments, we examine the task described by $\varphi_{reachAvoid}$ from Example \ref{ex: roToGo} and 
\begin{align*}
  \varphi_{stayIn} &= \square_{[0,20]}\code{region}, \text{ where} \\
  \code{region} &= (\bm{x}^r[0]-\bm{x}^e[0])^2 + (\bm{x}^r[1]-\bm{x}^e[1])^2 \!<\! 2.25,\\
\end{align*}
  which specifies that the robot must remain within some region for twenty seconds. The robot begins at $\bm{x}^r_0 = [0.5,2.5]^T$ for $\varphi_{reachAvoid}$ and $\bm{x}^r_0 = [1.5,2.5]^T$ for $\varphi_{stayIn}$. The environment state begins at $\bm{x}^e_0 = [3,2.5]^T$ for $\varphi_{reachAvoid}$ and $\bm{x}^e_0 = [2.5,2.5]^T$ for $\varphi_{stayIn}$. Fig. \ref{fig:bothEnvs} shows the environments for $\varphi_{reachAvoid}$ and $\varphi_{stayIn}$.

Our experiments investigate the impact of using \rotogotext as an objective function in an MPC algorithm. As baselines, we also use traditional robustness \cite{STLRobustness} and Arithmetic-Geometric Mean (AGM) Robustness \cite{smooth4}, one of the ``average'' robustness measures designed to address the locality issues discussed in Section \ref{sec:RelatedWork} \cite{daddyBelta}. 

Table \ref{tab:boxPlots} shows the averaged performance of all three approaches across 50 simulations, reporting the mean and standard deviation across a variety of measures. Along with robustness and AGM robustness, we report the success rate, trajectory length, and minimum distance from the dynamic environment. Our simulation stops upon formula violation, and because robustness, AGM robustness, and trajectory length are functions of complete trajectories, these values are representative of successful trials only. 


\begin{table*}
  \begin{center}
    \begin{tabular*}{\linewidth}{l@{\extracolsep{\fill}}lllll}
      \hline
      \textbf{Planning Problem} & 
      \textbf{Trajectory Length (m)} &
      \textbf{Min Distance (m)} &
      \textbf{Robustness} &
      \textbf{AGM Robustness} &
      \textbf{Success Rate (\%)} \\
      \hline
      $\varphi_{stayIn}$, Robustness & 5.112 $\pm$ 0.406 & -1.381 $\pm$ 0.211 & 0.238 $\pm$ 0.132 & 0.126 $\pm$ 0.014 & 80\\
      $\varphi_{stayIn}$, \rotogotext & 5.74 $\pm$ 0.522 & $\bm{-1.444 \pm 0.134}$ & $\bm{0.354 \pm 0.119}$ & $\bm{0.152 \pm 0.01}$ & $\bm{96}$\\
      $\varphi_{stayIn}$, AGM & 6.867 $\pm$ 0.604 & $\bm{-1.45 \pm 0.135}$ & $\bm{0.369 \pm 0.115}$ & $\bm{0.155 \pm 0.007}$ & $\bm{98}$\\
      \hline
      $\varphi_{reachAvoid}$, Robustness & 6.046 $\pm$ 0.375 & 0.236 $\pm$ 0.252 & $\bm{0.077 \pm 0.022}$ & 0.162 $\pm$ 0.012 & 68 \\
      $\varphi_{reachAvoid}$, \rotogotext & 7.246 $\pm$ 0.567 & 0.266 $\pm$ 0.267 & 0.069 $\pm$ 0.03 & 0.163 $\pm$ 0.012 & $\bm{74}$ \\
      $\varphi_{reachAvoid}$, AGM & 7.621 $\pm$ 0.514 & $\bm{0.404 \pm 0.474}$ & 0.046 $\pm$ 0.023 & $\bm{0.208 \pm 0.007}$ & 72\\
      \hline
    \end{tabular*}
  \end{center}
  \caption{Results of 50 runs of each approach against two STL specifications, simulated with a dynamic environment. The results for trajectory length, robustness, and AGM robustness represent successful trials only.}\label{tab:boxPlots}
\end{table*}

\subsection{Case Study 1: $\varphi_{stayIn}$}

  Fig. \ref{fig:stayIn} shows closed loop trajectories from all three algorithms when simulated in a \textit{static} environment. The circle represents the dynamic environment captured in \code{region} in $\varphi_{stayIn}$. Fig. \ref{fig:stayIn} shows that the robot stays near the center of the region after steering to it when using \rotogotext or AGM robustness, but steers away from the center when using traditional robustness. This is due to the proximity to the edge of the circle at the beginning of the mission. When using traditional robustness, the robot cannot distinguish between trajectories that stay near the middle and those that do not, because this proximity at the beginning limits the achievable robustness of plans.

  Table \ref{tab:boxPlots} shows the benchmarking results for this experiment, now simulated using a \textit{dynamic} environment. The minimum distance measures are negative because the robot is inside of the region, and the lowest attainable value is -1.5. The robot performs very similarly when using \rotogotext and AGM robustness in all measures except for trajectory length, where AGM robustness results in longer trajectories. Both AGM robustness and \rotogotext outperform traditional robustness.

\subsection{Case Study 2: $\varphi_{reachAvoid}$}


%
Fig. \ref{fig:reachAvoid} shows closed loop trajectories from all three algorithms when simulated in a static environment. The circle represents the human, described as subformula \code{human} in Example \ref{ex: roToGo}. Fig. \ref{fig:reachAvoid} shows that the robot steers further from the human when using \rotogotext as the objective function than it does when using traditional robustness. This is due to the proximity to the obstacles at the beginning of the mission. Online, when using traditional robustness, each plan is limited by this initial proximity, so trajectories that steer further from the human are no more robust than those that steer close to the human. Using \rotogotext, however, removes this dependence and allows the robot to steer further from the human. 

Fig. \ref{fig:reachAvoid} also shows a shortcoming of using AGM robustness as a cost function: because the function is not dominated by the most critical point, the robot can still achieve a high AGM robustness even when coming very close to collision. This is reflected in Fig. \ref{fig:reachAvoid} when the robot steers much closer to the walls at the beginning of the mission.

Table \ref{tab:boxPlots} shows that the robot achieves similar performance with regards to both robustness and AGM robustness whether it uses robustness or \rotogotext as the cost function. Despite this, the robot achieves a higher success rate when using \rotogotext, performing comparably to AGM robustness. The reason for this performance difference is more apparent when examining the trajectory lengths and minimum distances to the environment. The robot stays further from the dynamic region using \rotogotext when compared to traditional robustness, resulting in longer trajectories.

\subsection{Discussion}
  One advantage to using AGM robustness over traditional robustness and \rotogotext is that it results in trajectories that are temporally robust as well, e.g., the robot will spend more time in the \code{goal} region for $\varphi_{reachAvoid}$. This is in contrast to the behavior seen when using \rotogotext in Fig. \ref{fig:reachAvoid}: The robot does not go deeper into the goal nor stay in it, but instead leaves. This is because, once the robot has entered the goal region, the subformula \code{goal} in $\varphi_{reachAvoid}$ progresses to $\top$ and is not considered in the \rotogotext afterwards. One interpretation of this behavior is that the robot displays \textit{more} robust behavior under ``globally'' ($\square$) type formulas, and \textit{less} robust behavior under ``eventually'' ($\Diamond$) type formulas. In the context of MPC, however, there is a more pragmatic interpretation. STL robustness represents a trajectory's robustness to change. Because the past cannot change, satisfying points in the past are in some sense \textit{infinitely} robust (reflected in Definition \ref{def: rotogo}). There is no disturbance that would cause this past satisfaction to become a violation, and so planning to satisfy it more robustly can be wasteful.

  \rotogotext also has advantages over AGM robustness. Table \ref{tab:boxPlots} shows that the robot achieves a lower traditional robustness when using AGM robustness for $\varphi_{reachAvoid}$ when compared to \rotogotext. This is a result of the aforementioned behavior seen in Fig. \ref{fig:reachAvoid}: the robot can still have a high AGM robustness while coming close to collision at the beginning. \rotogotext is also less demanding to calculate online.
This is because, when leveraging formula progression for \rotogotext, the number of samples that must be considered decrements at each planning step while AGM robustness still depends on past states. 

  We reiterate that \rotogotext is an orthogonal concept to AGM, and a future direction for research could investigate the impact of using AGM robustness alongside formula progression.

\section{Conclusion and Future Work}
\label{sec:Conclusion}

The main contribution of this paper is STL robustness-to-go (\rotogotext), a new quantitative semantics for the logic, along with proofs of its relationship to MTL formula progression. This robustness measure addresses the tendency of events in the past limiting decision making capabilities of robotic systems when optimizing for traditional robustness. We implement robustness, robustness-to-go, and AGM robustness as the objective functions in an MPC algorithm, and find that robustness-to-go outperforms traditional robustness. 

For future work, we will investigate the relationship between robustness-to-go, which focuses on the contribution of \textit{suffix} trajectories, and the Robust Satisfaction Interval (RoSI) \cite{rosi}, which focuses on the contribution of \textit{prefix} trajectories. The work \cite{rcpe} uses the RoSI for MPC of unbounded STL specifications by considering the robustness of signals \textit{up to} a given point in time. RoSI as a cost function has similar issues to robustness, and combining it with robustness to go would enable the robot to consider only the robustness of a user-specified window of time. We plan to investigate how the RoSI-To-Go could address the locality issues and impact robot performance.

\bibliographystyle{IEEEtran}
\bibliography{references}

\appendices

  \section{Proofs}
  \label{sec: proofs}
  All proofs assume (i) STL is interpreted using pointwise semantics, (ii) time intervals $I$ are nonnegative, and (iii) sampling time $\Delta = t_{k+1} - t_k > 0$.
  \subsection{Theorem \ref{thm: RoToGoSound}}
\textit{A signal s satisfies an STL formula if and only if it has a positive robustness to go, i.e.
    \begin{equation*}
      (s,t) \vDash \varphi \iff \rotogo(s,t,\hat{t},\varphi) > 0.
    \end{equation*}}
    \proof{We provide a proof sketch. The result holds by induction on the structure of STL. The base case of $(s,t) \vDash \top \iff \rotogo(s,t,\hat{t},\top)$ is trivially true. The base case of $(s,t) \vDash g(\bm{x}) > 0 \iff \rotogo(s,t,\hat{t},g(\bm{x})>0)$ is true if and only if $g(s(t)) > 0$, regardless of whether $t>\hat{t}$. The proof for the inductive cases of negation, conjunction, and until follow from the analogous proof for traditional robustness, i.e., $(s,t) \vDash \varphi \iff \rho(s,t,\varphi) > 0$. This is because the \rotogotext of these operators according to Definition \ref{def: rotogo} are structurally identical to the robustness of these operators according to Definition \ref{def: robustness}. 
$\hfill \blacksquare$
  }

  \subsection{Theorem \ref{thm: progForRotogo}}
\begin{lemma}
When $\hat{t} < t_k$, robustness-to-go is equal to robustness, i.e., 
\begin{equation*}
  \hat{t} < t_k \rightarrow \rotogo(s,t_k,\hat{t},\varphi) = \rho(s,t_k,\varphi).
\end{equation*}
  \label{lem: roToGoRobEq}
\end{lemma}
  \proof{We provide a proof sketch. The result holds by induction on the structure of STL. The base case of $\hat{t} < t_k \rightarrow \rotogo(s,t_k,\hat{t},\top) = \rho(s,t_k,\top)$ is trivially true. The base case of $\hat{t} < t_k \rightarrow \rotogo(s,t_k,\hat{t},g(\bm{x}) > 0) = \rho(s,t_k,g(\bm{x}) > 0)$ is true according to Definitions \ref{def: robustness} and \ref{def: rotogo} because $\hat{t} < t_k$. The inductive cases for negation, conjunction, and until are straightforward because of the identical structure according to Definitions \ref{def: robustness} and \ref{def: rotogo}. For negation and conjunction, the inductive hypothesis holds because $\hat{t} < t_k$ for the child operators. The inductive hypothesis for the until operator also holds because $t',t'' > t_k$ in Definition \ref{def: rotogo} for nonnegative intervals $I$, so $\hat{t} < t_k \rightarrow \hat{t} < t', t''$. 
$\hfill \blacksquare$
  }

\begin{lemma}
When $\hat{t} = t_k$, robustness-to-go is equal to robustness of the progressed formula, i.e., 
\begin{equation*}
  \hat{t} = t_k \rightarrow \rotogo(s,t_k,\hat{t},\varphi) = \rho(s,t_{k+1},P(\varphi,t_{k+1}-t_k,s(t_k))).
\end{equation*}
  \label{lem: roToGoProgEq}
\end{lemma}

\proof{
  We provide a proof sketch. The result holds by induction on the structure of STL. The base case of $\hat{t} = t_k \rightarrow \rotogo(s,t_k,\hat{t},\top) = \rho(s,t_{k+1},P(\top, t_{k+1}-t_k, s(t_k)))$ is trivially true. According to Definitions \ref{def: formProg}, \ref{def: robustness}, and \ref{def: rotogo}, the base case of $\hat{t} = t_k \rightarrow \rotogo(s,t_k,\hat{t},g(\bm{x})>0) = \rho(s,t_{k+1},P(g(\bm{x})>0, t_{k+1}-t_k, s(t_k)))$ is true because both resolve to $\infty \cdot sign(g(s(t)))$ when $\hat{t} = t_k$. The inductive cases for negation and conjunction follow from the shared structure of these operations according to Definitions \ref{def: formProg}, \ref{def: robustness}, and \ref{def: rotogo}, and the inductive hypothesis holds because $\hat{t} < t_k$ for the child operators. The inductive case for the until operation relies on leveraging the pointwise semantics to separate the atomic contribution of the point $t_k$ from the \rotogotext, that is, 

  \begin{equation}
    \rotogo(s,t_k,\hat{t},\varphi U_I \psi) = 
    \begin{cases}
      \max(A,\min(B,C)) & \text{ if } 0 \in I \\
      \min(B,C) &\text{ otherwise,}
    \end{cases}
    \label{eq: checkpoint}
  \end{equation}
  where
  \begin{align}
    &A = \rotogo(s,t_k,\hat{t},\psi) \label{eq: dirtyA}\\
    &B = \rotogo(s,t_k,\hat{t},\varphi) \label{eq: dirtyB}\\
    &C = \rotogo(s,t_{k+1},\hat{t},\varphi U_{I^\leftarrow_{(t_{k+1} - t_k)}} \psi).\label{eq: dirtyC}
  \end{align}

  By the induction hypothesis, Equations \ref{eq: dirtyA} and \ref{eq: dirtyB} are equivalent to $\rho(s, t_{k+1}, P(\psi, t_{k+1}-t_k, s(t_k)))$ and $\rho(s, t_{k+1}, P(\varphi, t_{k+1}-t_k, s(t_k)))$, respectively, because $\hat{t} = t_k$, and Equation \ref{eq: dirtyC} becomes $\rho(s,t_{k+1},\varphi U_{I^\leftarrow_{(t_{k+1} - t_k)}} \psi)$ due to Lemma \ref{lem: roToGoRobEq} because $t_{k+1} > t_k$ and $t_k = \hat{t}$ imply $t_{k+1} > \hat{t}$. According to Definitions \ref{def: robustness} and \ref{def: formProg}, Equation \ref{eq: checkpoint} is therefore equivalent to $\rho(s,t_{k+1},P(\varphi U_I \psi, t_{k+1} - t_k, s(t_k)))$.
$\hfill \blacksquare$
  }

\begin{lemma}
When $\hat{t} > t_k$, robustness-to-go is equal to robustness-to-go of the progressed formula, i.e., 
  \begin{align*}
    \hat{t} > t_k \rightarrow \rotogo(s,t_k,&\hat{t},\varphi) \\ 
    &= \rotogo(s,t_{k+1},\hat{t},P(\varphi,t_{k+1}-t_k,s(t_k))).
  \end{align*}
  \label{lem: rotoGoOneStep}
\end{lemma}

\proof{We provide a proof sketch. The result holds by induction on the structure of STL. The base case of $\hat{t} > t_k \rightarrow \rotogo(s,t_k,\hat{t},\top) = \rotogo(s,t_{k+1},\hat{t},P(\top, t_{k+1}-t_k, s(t_k)))$ is trivially true. The base case of $\hat{t} > t_k \rightarrow \rotogo(s,t_k,\hat{t},\top) = \rotogo(s,t_{k+1},\hat{t},P(\top, t_{k+1}-t_k, s(t_k)))$ follows from Definitions \ref{def: formProg} and \ref{def: rotogo} for the case that $\hat{t} > t_k$. The inductive cases for negation and conjunction follow from the shared structure of these operations according to Definitions \ref{def: formProg} and \ref{def: rotogo}, and the inductive hypothesis holds because $\hat{t} > t_k$ for the child operators. Similar to the proof of Lemma \ref{lem: roToGoProgEq}, the inductive case for the until operation relies on separating the atomic contribution of the point $t_k$ from the \rotogotext, resulting in Equation \ref{eq: checkpoint}. By inductive hypothesis, Equations \ref{eq: dirtyA} and \ref{eq: dirtyB} are equivalent to $\rotogo(s, t_{k+1}, \hat{t}, P(\psi, t_{k+1} - t_k, s(t_k)))$ and $\rotogo(s, t_{k+1}, \hat{t}, P(\varphi, t_{k+1} - t_k, s(t_k)))$, respectively, because $\hat{t} > t_k$. According to Definitions \ref{def: formProg} and \ref{def: rotogo}, Equation \ref{eq: checkpoint} is therefore equivalent to $\rotogo(s,t_{k+1},\hat{t},P(\varphi U_I \psi,t_{k+1}-t_k,s(t_k)))$.
$\hfill \blacksquare$
}

\begin{lemma}
Let $\varphi_0$ be an STL formula, $s$ be a signal, and consider sample $(t_i,s(t_i))$. Let $\Phi = \{\varphi_0, \varphi, ..., \varphi_{i+1}\}$ be a set of STL formulas such that $\varphi_{k} = P(\varphi_{k-1},t_k-t_{k-1},s(t_{k-1}))$ for each $0 < k \leq i+1$. Then, 
\begin{equation*}
  t_0 < t_k \leq t_i \rightarrow \rotogo(s,t_k,t_i,\varphi_k) = \rotogo(s,t_0,t_i,\varphi_0).
\end{equation*}
  \label{lem: theoremHelperProof}
\end{lemma}
\proof{We provide a proof sketch. The result holds by induction of Lemma \ref{lem: rotoGoOneStep} through time. Because $t_k \leq t_i$ and $t_{k-1} < t_k$, we know $t_{k-1} < t_i$. From Lemma \ref{lem: rotoGoOneStep}, this means that $\rotogo(s,t_k,t_i,\varphi_k) = \rotogo(s,t_{k-1},t_i,\varphi_{k-1})$. The base case is $t_1 = t_k \leq t_i$, meaning $\rotogo(s,t_k,t_i,\varphi_k) = \rotogo(s,t_0, t_i,\varphi_0)$. In the inductive case, $t_1 < t_k \leq t_i$, implying that $t_0 < t_{k-1} < t_i$. This allows for application of the inductive hypothesis, meaning $\rotogo(s,t_{k-1},t_i,\varphi_{k-1}) = \rotogo(s,t_0,t_i,\varphi_0)$. Accordingly, $\rotogo(s, t_k, t_i, \varphi_k) =\rotogo(s,t_{k-1},t_i,\varphi_{k-1})= \rotogo(s,t_0,t_i,\varphi_0)$ for all $0 < k \leq i$. 
$\hfill \blacksquare$
}

  We now turn our attention to Theorem \ref{thm: progForRotogo}:

Let $\varphi_0$ be an STL formula, $s$ be a signal, and consider sample $(t_i,s(t_i))$. Let $\Phi = \{\varphi_0, \varphi, ..., \varphi_{i+1}\}$ be a set of STL formulas such that $\varphi_{k} = P(\varphi_{k-1},t_k-t_{k-1},s(t_{k-1}))$ for each $0 < k \leq i+1$. Then, 
\begin{equation*}
  t_i \geq t_0 \rightarrow \rho(s,t_{i+1},\varphi_{i+1}) = \rotogo(s,t_0,t_i,\varphi_0).
\end{equation*}

\proof{From Lemma \ref{lem: roToGoProgEq}, $\rho(s,t_{i+1},\varphi_{i+1}) = \rotogo(s,t_0,t_i,\varphi_0)$ when $t_i = t_0$. 
From Lemma \ref{lem: theoremHelperProof}, $\rotogo(s,t_0,t_i,\varphi_0) = \rotogo(s,t_k,t_i,\varphi_k)$ for any $t_0 < t_k \leq t_i$ when $t_i > t_0$. If we chose $t_k = t_i$,
$
  \rotogo(s,t_0,t_i,\varphi_0) = \rotogo(s,t_i,t_i,\varphi_i),
$
and, from Lemma \ref{lem: roToGoProgEq},
$
\rotogo(s,t_i,t_i,\varphi_i) = \rho(s,t_{i+1},\varphi_{i+1}).
$
$\hfill \blacksquare$
} 

\section{Hyperparameters}
\label{sec: params}

\begin{table}[h]
  \centering
    \begin{tabular*}{\linewidth}{l@{\extracolsep{\fill}}ll}
      \hline
      \textbf{Parameter} & 
      \textbf{Value} \\
      \hline
      VP-STO degrees of freedom & 2\\
      VP-STO velocity limits & 2 $m/s$\\
      VP-STO acceleration limits & 2 $m/s^2$\\
      VP-STO number of via points & 4\\
      VP-STO population size & 10\\ 
      VP-STO CMA-ES initial variance & 10 $m^2$\\
      VP-STO maximum CMA-ES iterations & 50\\
      X domain lower bound & 0 $m$\\
      X domain upper bound & 5 $m$\\
      Y domain lower bound & 0 $m$\\
      Y domain upper bound & 5 $m$\\
      Cost penalty for violating domain bounds & 1e8\\
      Planning Horizon & 20 $s$\\
      \hline
    \end{tabular*}
  \caption{Experiment hyperparameters.}\label{tab:params}
\end{table}

\end{document}